\documentclass[aps,prl,reprint,showpacs,byrevtex]{revtex4-2}
\let\oldhat\hat
\renewcommand{\hat}[1]{\oldhat{\mathbf{#1}}}
\usepackage{titlesec}
\usepackage{array}

\usepackage{times,mathptm}
\usepackage[dvips]{graphicx,color}

\usepackage{amsmath}
\usepackage{xcolor}

\begin{document}
\title{Microscopic origin of the magnetic easy-axis switching in Fe$_3$GaTe$_2$ under pressure}
\author{Jiaqi Li$^{1}$, Shuyuan Liu$^{2}$, Chongze Wang$^{1,2}$, Fengzhu Ren$^1$, Bing Wang$^{1*}$, and Jun-Hyung Cho$^{1,2*}$}
\affiliation{$^1$Joint Center for Theoretical Physics, School of Physics and Electronics, Henan University, Kaifeng 475004, China \\
$^2$Department of Physics and Research Institute for Natural Science, Hanyang University, 222 Wangsimni-ro, Seongdong-Ku, Seoul 04763, Repu\emph{}blic of Korea}
\date{\today}

\begin{abstract}
The two-dimensional layered ferromagnet Fe$_3$GaTe$_2$, composed of a Te–Fe$_{\rm I}$–Fe$_{\rm II}$/Ga–Fe$_{\rm I}$–Te stacking sequence, hosts two inequivalent Fe sites and exhibits a high Curie temperature and strong out-of-plane magnetic anisotropy, making it a promising platform for spintronic applications. Recent experiments have observed a pressure-induced switching of the magnetic easy axis from out-of-plane to in-plane near 10 GPa, though its microscopic origin remains unclear. Here, we employ first-principles calculations to investigate the pressure dependence of the magnetocrystalline anisotropy energy in Fe$_3$GaTe$_2$. Our results reveal a clear easy-axis switching at a critical pressure of approximately 10~GPa, accompanied by a sharp decrease in the magnetic moments arising from Fe$_{\rm I}$ and Fe$_{\rm II}$ atoms. As pressure increases, spin-up and spin-down bands broaden and shift oppositely due to band dispersion effects, leading to a reduction in net magnetization. Simultaneously, the SOC contribution from Fe$_{\rm I}$, which initially favors an out-of-plane easy axis, diminishes and ultimately changes sign, thereby promoting in-plane anisotropy. The SOC contribution from the outer-layer Te atoms also decreases steadily with pressure, although it retains its original sign; this additional reduction further reinforces the in-plane magnetic easy axis. In contrast, Fe$_{\rm II}$ atoms continue to favor an out-of-plane orientation, but their contribution is insufficient to counterbalance the dominant in-plane preference at high pressure. These findings elucidate the origin of magnetic easy-axis switching in Fe$_3$GaTe$_2$ and provide insights for tuning magnetic anisotropy in layered materials for spintronic applications.
\end{abstract}

\maketitle

\section{I. INTRODUCTION}

In recent years, two-dimensional (2D) van der Waals (vdW) magnetic materials have attracted growing interest in spintronics owing to their intrinsic spin-dependent properties and highly tunable magnetic ordering. The discovery of intrinsic ferromagnetism in several 2D vdW compounds, including CrI$_3$~\cite{Nature-CrI3}, Cr$_2$Ge$_2$Te$_6$~\cite{Nature-Cr2Ge2Te6}, Fe$_3$GeTe$_2$~\cite{Nature Materials-Fe3GeTe2}, Fe$_5$GeTe$_2$~\cite{npj-Fe5GeTe2}, Fe$_3$GaTe$_2$~\cite{NC2022-Fe3GaTe2}, VSe$_2$~\cite{Nature Nanotechnology-VSe2}, MnSe$_2$~\cite{NanoLetters-MnSe2}, CrSe$_2$~\cite{Nature Materials-CrSe2} and CrTe$_2$~\cite{NC-CrTe2} has opened up new opportunities for exploring low-dimensional magnetism and associated spin transport phenomena. Unlike their three-dimensional counterparts, 2D vdW magnets allow for effective control of magnetic properties through external stimuli such as carrier doping, layer thickness, mechanical strain, and hydrostatic pressure. This high tunability makes them ideal platforms for next-generation spintronic, spin-caloritronic, magneto-optical, and quantum information devices, where precise modulation of magnetic states is crucial.~\cite{Nature-Fe3GeTe2,NC-Fe4GeTe2}

Recently, Zhang \emph{et al}.~\cite{NC2022-Fe3GaTe2} successfully synthesized high-quality Fe$_3$GaTe$_2$ single crystals via the self-flux method and reported a record-high Curie temperature ($T_c$) of approximately 350--380~K among vdW layered ferromagnets, significantly surpassing that of the widely studied sister compound Fe$_3$GeTe$_2$~\cite{Nature Materials-Fe3GeTe2}. Beyond its elevated $T_c$, Fe$_3$GaTe$_2$ exhibits remarkable magnetic characteristics, including strong perpendicular magnetocrystalline anisotropy, a large saturation magnetic moment, and a sizable anomalous Hall angle, all of which persist above room temperature. Among these properties, magnetic anisotropy is particularly important for information storage technologies, as it enhances thermal stability, increases data density, and reduces power consumption, which are key factors for the development of next-generation spintronic and magnetic memory devices~\cite{APL-PMA,JACS-SMM-MOFs}. More recently, pressure-dependent measurements have revealed a sharp reorientation of the magnetic anisotropy in Fe$_3$GaTe$_2$, where the magnetic easy axis switches from out-of-plane to in-plane within the pressure range of 10.3--12.2~GPa~\cite{Advanced Science-Fe3GaTe2 pressure}. This transition has been tentatively attributed to a competition between interlayer and intralayer exchange interactions, along with an increased density of states near the Fermi level ($E_F$) under pressure~\cite{Advanced Science-Fe3GaTe2 pressure}. However, the microscopic mechanism underlying this anisotropy switching remains unresolved. Understanding the origin of this pressure-induced transition is crucial for elucidating the interplay among lattice structure, electronic states, and spin-orbit coupling (SOC). Such insights are essential not only for advancing fundamental knowledge but also for enabling the rational design of 2D magnetic materials with tunable anisotropy. In this study, we address this issue by performing systematic first-principles calculations to uncover the microscopic origin of the magnetic anisotropy transition in Fe$_3$GaTe$_2$ under pressure.

In this work, we perform first-principles density functional theory (DFT) calculations to investigate the evolution of the magnetic anisotropy energy (MAE) in Fe$_3$GaTe$_2$ under hydrostatic pressure ranging from 0 to 20 GPa. Our calculations reveal a clear spin reorientation transition of the magnetic easy axis from out-of-plane to in-plane at a critical pressure of approximately 10 GPa, consistent with experimental observations~\cite{Advanced Science-Fe3GaTe2 pressure}. As pressure increases, we also find electronic band broadening, which leads to opposite shifts in the spin-up and spin-down bands and a marked decrease in magnetization near 10 GPa. At the same time, SOC contributions from inequivalent atomic sites exhibit distinct pressure-dependent behavior. For Fe$_{\rm I}$ atoms in the outer layers, the initially out-of-plane–favoring anisotropy weakens with increasing pressure and eventually reverses to support in-plane alignment. Te atoms at the top and bottom of the vdW structure exhibit a similar decreasing trend in their SOC effect, though without a sign change, thereby further reinforcing the in-plane magnetic preference. In contrast, Fe$_{\rm II}$ atoms in the central layer retain a weak out-of-plane preference, but their contribution remains too small to counterbalance the dominant in-plane-driving effects from Fe$_{\rm I}$ and Te at high pressure. These atom-resolved insights clarify the microscopic origin of the pressure-induced magnetic anisotropy transition in Fe$_3$GaTe$_2$, highlighting the interplay between electronic band evolution and site-specific SOC effects. Our findings demonstrate that pressure offers an effective route to modulate magnetic anisotropy in 2D vdW magnets, providing useful guidance for the design of spintronic devices with controllable magnetic functionality.

\section{II. CALCULATIONAL METHODS}

Our first-principles DFT calculations were performed using the Vienna \textit{ab initio} simulation package (VASP)~\cite{vasp1,vasp2} with a plane-wave basis set. The interactions between core and valence electrons were treated using the projector augmented-wave (PAW) method~\cite{paw}. Spin-polarized local density approximation (LDA)~\cite{lda} with the Perdew–Zunger parameterization of the Ceperley–Alder (CA) Monte Carlo correlation data was used to approximate the exchange-correlation functional. To maintain consistency, the PAW potentials were generated using the same CA-LDA functional. A kinetic energy cutoff of 500~eV was applied, and the electronic self-consistent energy convergence criterion was set to $10^{-8}$~eV. Structural optimizations were carried out until the Hellmann–Feynman forces on each atom were less than 0.001~eV/\AA. The Brillouin zone was sampled using a $18 \times 18 \times 5$ $k$-point mesh. To accurately capture vdW interactions, the DFT-D3 dispersion correction scheme was employed~\cite{JCP154104,JCC1456}.

\section{III. RESULTS and DISCUSSION}

We begin by optimizing the crystal structure of Fe$_3$GaTe$_2$ under hydrostatic pressure using first-principles DFT calculations~\cite{PRESSURE}. Fe$_3$GaTe$_2$ crystallizes in a layered hexagonal structure with space group $P6_3/mmc$ (No.~194). As illustrated in Figs.~1(a) and 1(b), the conventional unit cell contains two formula units (f.u.) and comprises two weakly coupled slabs stacked along the crystallographic $c$-axis, held together by vdW interactions. Each slab consists of five atomic sublayers: Te atoms occupy the top and bottom layers, Fe$_{\rm I}$ atoms are situated in the second and fourth layers, and the central layer hosts both Fe$_{\rm II}$ and Ga atoms. Within each sublayer, the atoms form 2D triangular lattices. In particular, three Te atoms at the corners of a triangle bond with a single Fe$_{\rm I}$ atom positioned at the triangle's hollow site, forming a characteristic local coordination geometry. Meanwhile, Ga atoms occupy the in-plane hollow sites coordinated by the Fe$_{\rm II}$ sublattice. This arrangement results in a specific trilayered Fe$_3$Ga framework sandwiched between two Te layers, yielding a quasi-2D structure. Upon full structural relaxation, the optimized lattice constants at zero pressure are found to be $a = b = 3.901$~\AA{} and $c = 15.221$~\AA{}, with an interlayer spacing of $d_{\text{int}} = 2.677$~\AA{} (see Table I). Figure~1(c) shows the pressure dependence of the lattice parameters $a$, $b$, and $c$, as well as $d_{\text{int}}$. While all structural parameters decrease monotonically with increasing pressure, the compressibility along the $c$-axis is significantly greater than in the in-plane directions, resulting in a more pronounced reduction in $c$ and $d_{\text{int}}$ compared to $a$ and $b$. This anisotropic structural response reflects the intrinsic softness of the vdW gap under pressure.
\begin{figure}
    \centering
    \includegraphics[width=1\linewidth]{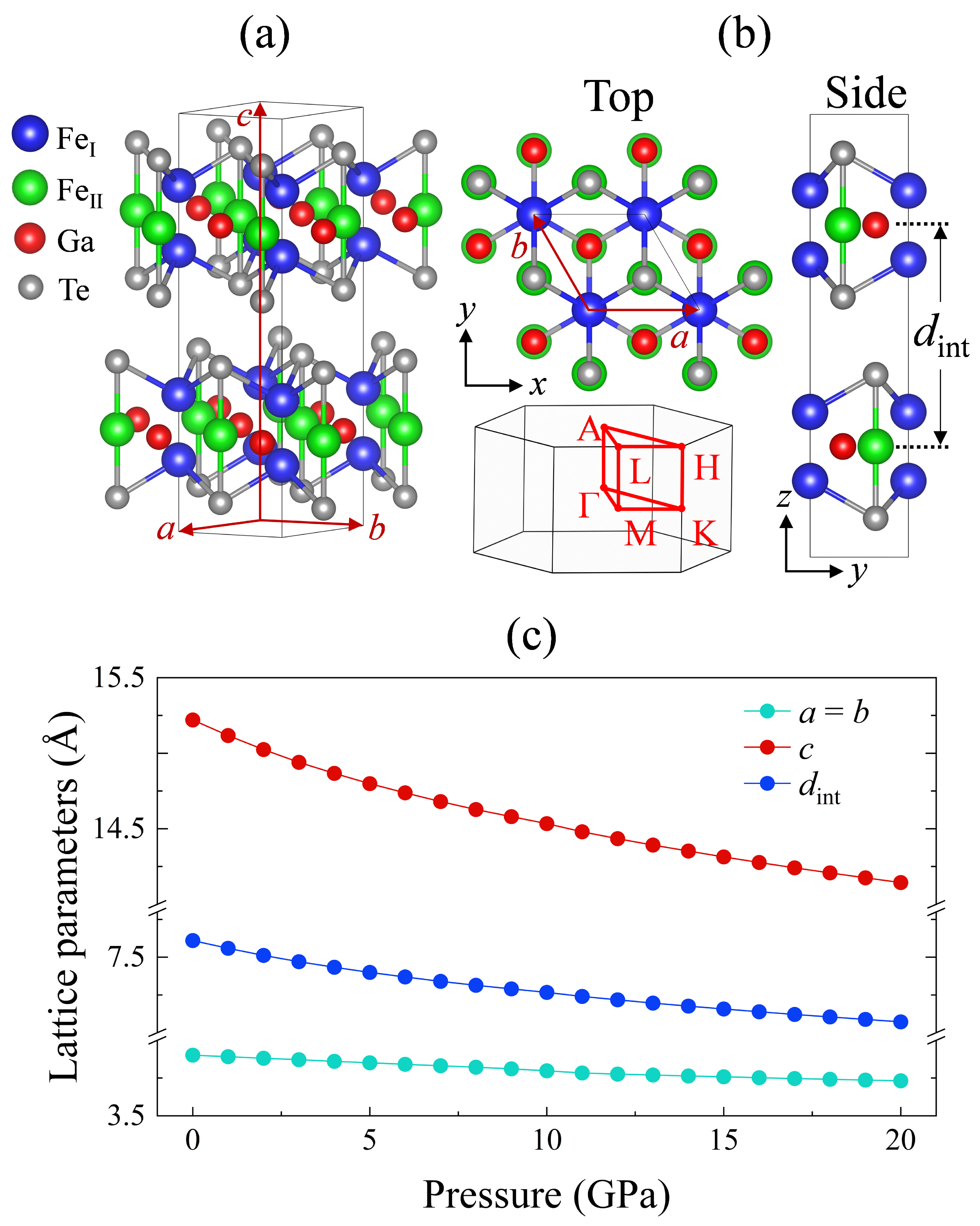}
    \caption{Optimized crystal structure of Fe$_3$GaTe$_2$: (a) Perspective view of the layered structure; (b) top and side views of the unit cell, together with the first Brillouin zone of the hexagonal lattice; (c) calculated pressure dependence of the lattice parameters $a$, $b$, and $c$, as well as the interlayer spacing $d_{\text{int}}$. The lattice parameters are $a$ = ($a$, 0, 0), $b$ = ($-b$ sin(30$^{\circ}$), $b$ cos(30$^{\circ}$), 0), and $c$ = (0, 0, $c$) in the $x$, $y$, $z$ coordinate system.}
    \label{fig:enter-label}
\end{figure}

\begin{table}[ht]
\caption{Calculated lattice parameters $a$, $b$, $c$, and interlayer spacing $d_{\text{int}}$ of Fe$_3$GaTe$_2$ at zero pressure~\cite{CALDA}, in comparison with previous theoretical~\cite{PRB-Xu} and experimental~\cite{CommunMater-Fe3GaTe2,Nano Letters-Fe3GaTe2,NC2022-Fe3GaTe2} results. The employed exchange-correlation functionals, including CA-LDA and the generalized gradient approximation functional of Perdew, Burke, and Ernzerhof (PBE-GGA)~\cite{PBE-GGA}, are indicated in parentheses. }
\begin{ruledtabular}
\begin{tabular}{lcccc}
       &  $a$ ({\AA}) & $b$ ({\AA}) & $c$ ({\AA}) & $d_{\text{int}}$ ({\AA}) \\ \hline
Present (CA-LDA) & 3.901 & 3.901  & 15.221 & 2.677 \\
Present (PBE-GGA) & 4.032 & 4.019  & 16.109 & 2.887 \\
Previous theory (PBE-GGA)~\cite{PRB-Xu} & 4.034 & 4.034 & 16.090 \\
Experiment~\cite{NC2022-Fe3GaTe2} & 3.986  & 3.986  & 16.229  &   \\
Experiment~\cite{CommunMater-Fe3GaTe2} & 3.970 & 3.970 & 15.560 &  \\
Experiment~\cite{Nano Letters-Fe3GaTe2} & 4.090 & 4.090  & 16.070 & \\
\end{tabular}
\end{ruledtabular}
\end{table}

Figure~2(a) shows the spin-polarized band structure and local density of states (LDOS) of ferromagnetic Fe$_3$GaTe$_2$ at zero pressure, calculated using the LDA functional without SOC. The electronic states near $E_F$ are primarily derived from the Fe 3$d$ orbitals of both Fe$_{\rm I}$ and Fe$_{\rm II}$. The band structure exhibits hole-like dispersions along the $\Gamma$--$A$ direction and electron-like bands along $K$--$H$. The overall dispersions along high-symmetry paths such as $\Gamma$--$M$, $\Gamma$--$K$, $A$--$L$, and $A$--$H$ are consistent with previous DFT study and angle-resolved photoemission spectroscopy (ARPES) measurements~\cite{Nano Letters-Fe3GaTe2,CPL-Fe3GaTe2}. Figure~2(b) displays the calculated Fermi surface (FS) contours at $k_z = 0$ and $k_z = \pi/c$. At $k_z = 0$, the FS contains multiple hole pockets centered at $\Gamma$, including outer hexagonal contours, along with several triangular-like electron pockets at the $K$ points, reflecting the threefold rotational symmetry of the lattice. In contrast, at $k_z = \pi/c$, the FS shows moderate reconstruction: five hole pockets encircle $A$, and two triangular-like electron pockets appear at the $H$ points. The FS variation at $k_z = \pi/c$ arises from the non-symmorphic nature of the $P6_3/mmc$ space group, which enforces double degeneracy of all bands at $k_z = \pi/c$ in each spin channel. The band dispersion along the $\Gamma$--$A$ direction, as shown in Fig.~2(a), reflects finite interlayer dispersion along the c-axis, which originates from weak vdW interactions between adjacent layers. Notably, both occupied and unoccupied states at the $M$ and $L$ points give rise to van Hove singularities (VHSs) near $E_F$, resulting in prominent LDOS peaks. In particular, the occupied states generate VHS1 and VHS2 at the $M$ point, located at $-0.318$ and $-0.448$ eV below $E_F$, and VHS3 at the $L$ point, located at $-0.386$ eV [see Fig.~2(a)]. These features align well with the strong ARPES intensity distributed between $-0.3$ and $-0.5$ eV at the same momenta~\cite{Nano Letters-Fe3GaTe2}. Therefore, the LDA-derived electronic structure reliably captures both the FS topology and VHS-related LDOS features, providing a solid foundation for analyzing the pressure-induced evolution of electronic and magnetic properties in Fe$_3$GaTe$_2$.

\begin{figure}
    \centering
   \includegraphics[width=1\linewidth]{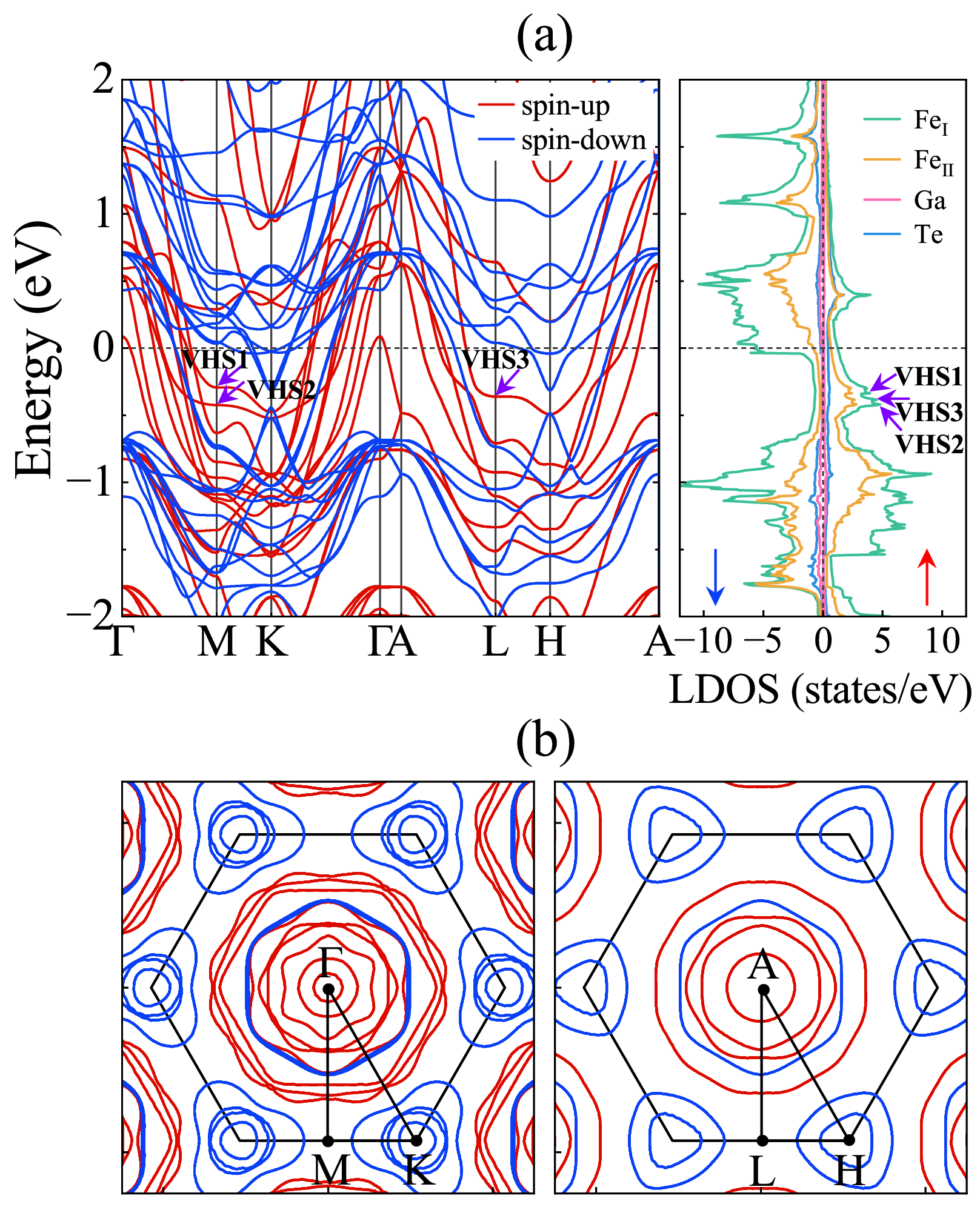}
    \caption{(a) Calculated spin-polarized band structure of Fe$_3$GaTe$_2$ at zero pressure, along with the LDOS projected onto the Fe$_{\rm I}$, Fe$_{\rm II}$, and Te atoms. The VHS1, VHS2, and VHS3 are marked in the band structure and LDOS in panel (a). The red and blue arrows represent the spin-up and spin-down LDOS, respectively. The LDOS is given in units of states/eV per f.u. (b) Corresponding FS contours in the $k_z = 0$ (left) and $k_z = \pi/c$ (right) planes. }
    \label{fig:enter-label}
\end{figure}

\begin{figure*}
    \centering
    \includegraphics[width=1\linewidth]{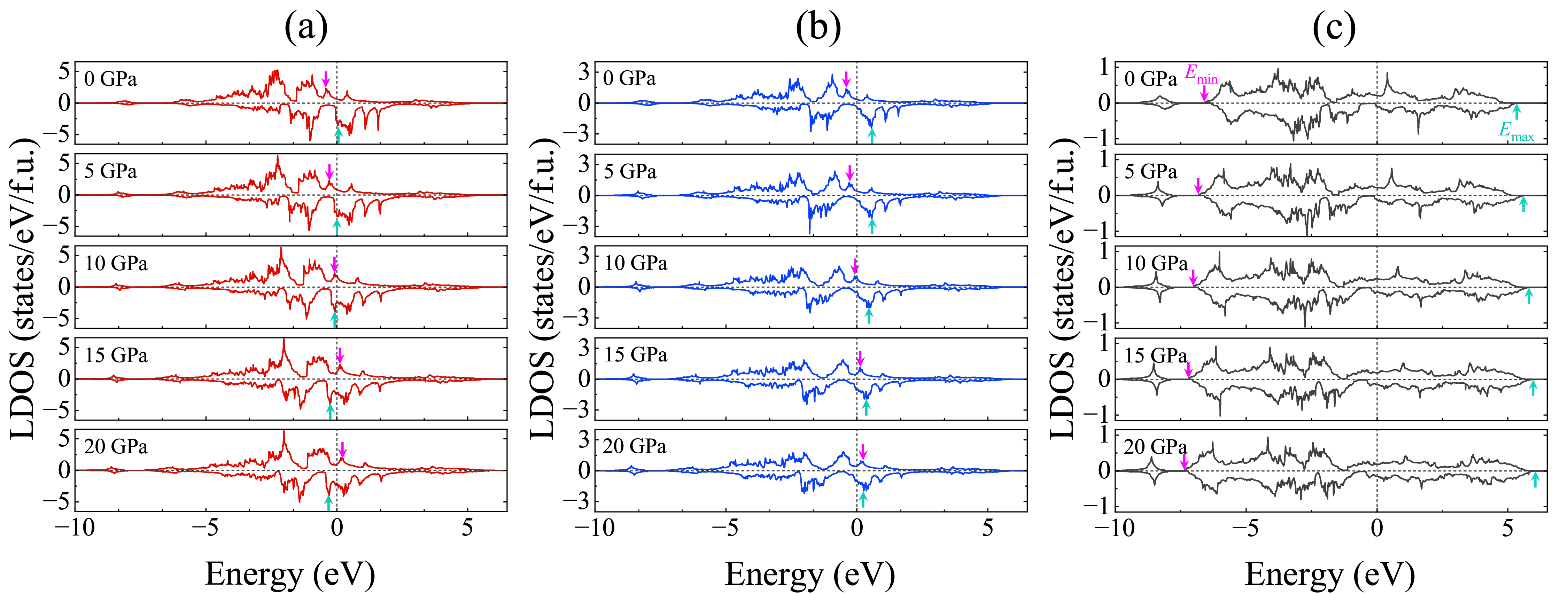}
    \caption{Calculated pressure-dependent LDOS for (a) Fe$_{\rm I}$, (b) Fe$_{\rm II}$, and (c) Te atoms. The arrows in panels (a) and (b) highlight the LDOS peaks near $E_F$. The maroon (cyan) arrows for the spin-up (spin-down) states in panels (a) and (b) highlight the LDOS peaks near $E_F$. In panel (c), the arrows indicate the lower ($E_{\rm min}$) and upper ($E_{\rm max}$) band edges, corresponding to the occupied and unoccupied states, respectively.}
    \label{fig:enter-label}
\end{figure*}

To investigate how the electronic structure evolves under pressure, we present the pressure-dependent LDOS of Fe$_{\rm I}$, Fe$_{\rm II}$, and Te atoms over a broad energy range [see Figs.~3(a)--3(c)]. With increasing pressure, all atomic species exhibit noticeable band broadening, indicating enhanced orbital hybridization. Notably, the LDOS of Te shows a systematic expansion of the band edges, denoted by $E_{\rm min}$ and $E_{\rm max}$, with pressure [see Fig. 3(c)]. The values of $E_{\rm min}$ ($E_{\rm max}$) are $-6.586$ ($5.690$), $-6.818$ ($5.874$), $-7.000$ ($6.018$), $-7.188$ ($6.147$), and $-7.355$ ($6.275$) eV at 0, 5, 10, 15, and 20~GPa, respectively. The LDOS peaks of Fe$_{\rm I}$ and Fe$_{\rm II}$ appear at similar energies, suggesting strong hybridization between the two Fe sublattices, likely mediated by Te $p$ orbitals. In addition to the overall broadening, the LDOS near $E_F$ exhibits spin-dependent energy shifts with increasing pressure [see arrows in Figs.~3(a) and 3(b)]: spin-up states shift upward, while spin-down states shift downward. This behavior reduces the exchange splitting and causes a redistribution of spectral weight from spin-up to spin-down channels, leading to a net suppression of spin polarization.

To clarify the orbital character of the states indicated by the arrows in Fig.~3, we have included the partial density of states (PDOS) for Fe$_{\rm I}$, Fe$_{\rm II}$, Te, and Ga atoms near $E_F$ in Supplemental Fig.~S2~\cite{SM}. For the spin-up states (indicated by maroon arrows), the dominant contributions come from the Fe$_{\rm I}$ $d_{z^2}$ orbital and the Fe$_{\rm II}$ $d_{xz}$/$d_{yz}$ orbitals. In contrast, the spin-down states (indicated by cyan arrows) involve Fe$_{\rm I}$ $d_{xz}$/$d_{yz}$ and Fe$_{\rm II}$ $d_{z^2}$ orbitals. The comparable spectral weight of the spin-up states on Fe$_{\rm I}$ and Fe$_{\rm II}$ suggests strong hybridization between Fe$_{\rm I}$ $d_{z^2}$ and Fe$_{\rm II}$ $d_{xz}$/$d_{yz}$ orbitals. For the spin-down states, the corresponding PDOS peaks in Te/Ga $p$ orbitals indicate hybridizations between Fe$_{\rm I}$ $d_{xz}$/$d_{yz}$ and Te/Ga $p_{z}$ orbitals, as well as between Fe$_{\rm II}$ $d_{z^2}$ and Ga $p_{x}$/$p_{y}$ orbitals. These different orbital hybridizations between Fe and Te/Ga atoms, where the Fe$_{\rm I}$/Te(Ga) interaction involves $p_{z}$ (out-of-plane) orbitals and the Fe$_{\rm II}$/Ga interaction involves $p_{x}$/$p_{y}$ (in-plane) orbitals, may be associated with the divergent pressure responses of the MAE for Fe$_{\rm I}$/Te and Fe$_{\rm II}$, as discussed below.

Figure~4 displays the pressure dependence of the magnetic moments for Fe$_{\rm I}$ and Fe$_{\rm II}$. At zero pressure, Fe$_{\rm I}$ has a larger magnetic moment than Fe$_{\rm II}$ possibly due to a more prominent unoccupied spin-down LDOS just above $E_F$, as shown in Fig.~2. The magnetic moment on Te remains negligible ($<{\sim}0.05~\mu_B$), indicating that magnetism is primarily localized on the Fe sublattices. As pressure increases up to 10~GPa, both Fe sites show a pronounced reduction in magnetic moment (see Fig. 4). For Fe$_{\rm I}$, this is primarily due to the downward shift of spin-down states located just above $E_F$, which become partially occupied as they move below the Fermi level. Additionally, spin-up states near $E_F$ shift upward and become depopulated, further contributing to the decline in spin polarization. A similar mechanism applies to Fe$_{\rm II}$, resulting in a simultaneous drop in moment. Beyond 10~GPa, the rate of moment reduction slows markedly, and the magnetic moments of Fe$_{\rm I}$ and Fe$_{\rm II}$ gradually converge. This crossover reflects more subtle, nonuniform changes in the band structure near $E_F$, and marks the onset of saturation in spin-state redistribution. Notably, this regime coincides with the emergence of magnetic anisotropy switching, which will be discussed below. The numbers in Fig.~4 show the spin-resolved Fe $d$-electron counts for both Fe$_{\rm I}$ and Fe$_{\rm II}$ sites as a function of pressure, illustrating how the $d$-orbital occupations evolve and correlate with the pressure-induced reduction in magnetic moments. Collectively, these results underscore the itinerant nature of magnetism in Fe$_3$GaTe$_2$. The continuous and pressure-tunable variation of magnetic moments, driven by changes in the electronic structure near $E_F$, is a characteristic feature of itinerant ferromagnetism in layered vdW materials.

\begin{figure}
    \centering
    \includegraphics[width=1\linewidth]{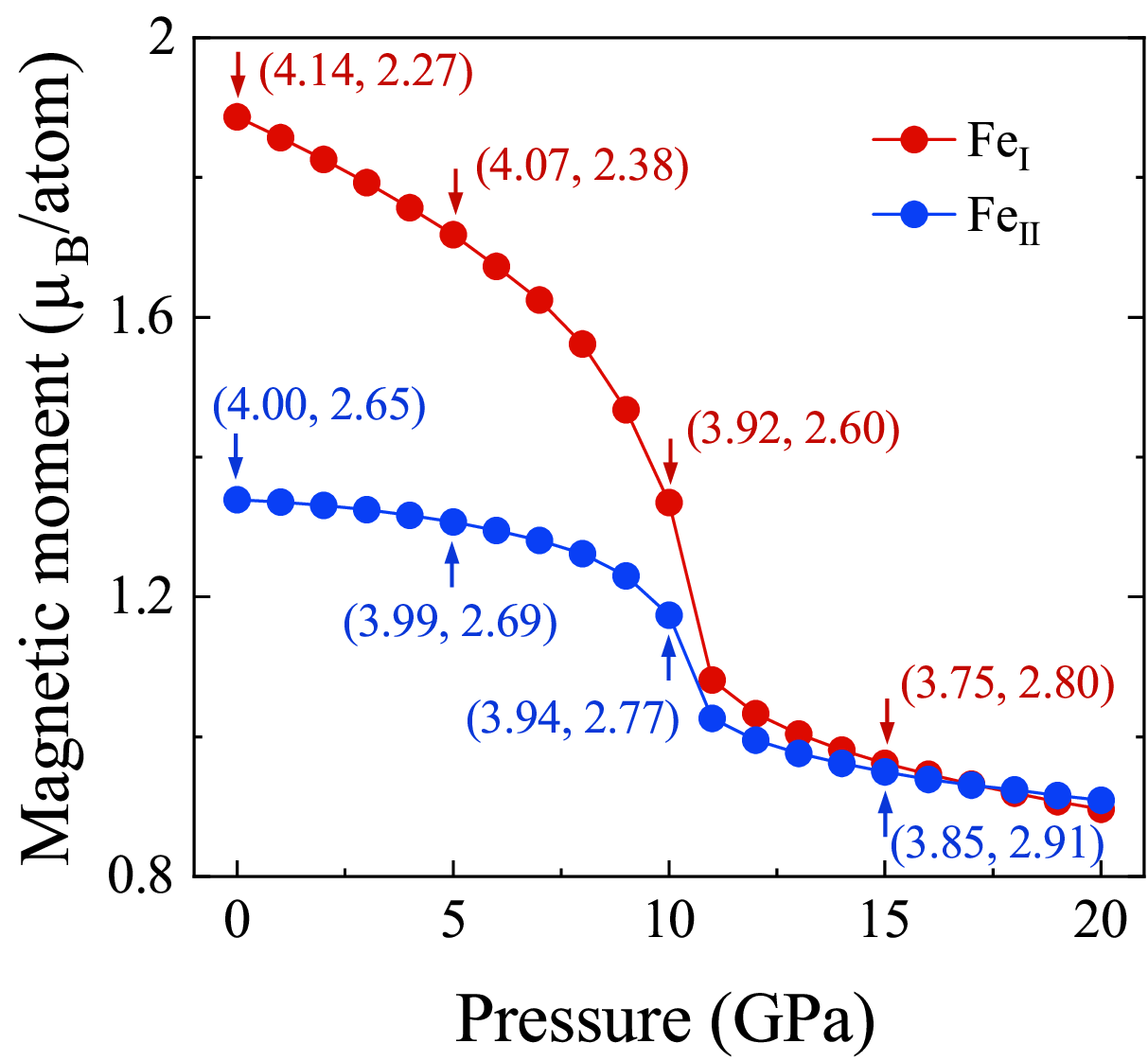}
    \caption{Calculated pressure dependence of the spin magnetic moments of Fe$_{\rm I}$ and Fe$_{\rm II}$. The paired numbers in parentheses represent the spin-up and spin-down Fe $d$-electron counts, respectively.}
    \label{fig:enter-label}
\end{figure}

Next, we investigate the MAE of Fe$_3$GaTe$_2$ by including SOC~\cite{MAE-SOC}. The MAE is defined as the total energy difference between in-plane and out-of-plane magnetization directions, i.e., ${\rm MAE} = E_{\parallel} - E_{\perp}$, where $E_{\parallel}$ and $E_{\perp}$ denote the total energies for in-plane and out-of-plane magnetizations, respectively. In this convention, a positive MAE indicates an out-of-plane magnetic easy axis, while a negative value corresponds to an in-plane one. As shown in Fig.~5(a), the MAE at zero pressure is positive, with a value of 0.677 meV/f.u., indicating that the easy axis lies along the $c$-axis. Figure~5(b) shows the pressure dependence of the MAE. With increasing pressure, the MAE varies weakly before decreasing, and ultimately changes sign near 10~GPa, indicating a pressure-driven spin reorientation from an out-of-plane to an in-plane easy axis. This behavior is consistent with recent experimental observations~\cite{Advanced Science-Fe3GaTe2 pressure}. Notably, this critical pressure for MAE sign reversal coincides with a pronounced drop in the magnetic moments of Fe$_{\rm I}$ and Fe$_{\rm II}$ (see Fig.~4), reflecting a common microscopic origin. The reorientation is likely driven by pressure-induced modifications in the spin-resolved electronic states near $E_F$: occupied spin-up bands shift upward, while unoccupied spin-down bands move downward (see Fig.~3). These opposing shifts lead to high LDOS peaks associated with Fe$_{\rm I}$ and Fe$_{\rm II}$ crossing $E_F$, resulting in a redistribution of spin-resolved occupations. The subsequent reduction of local magnetic moments alters the SOC interactions, ultimately reversing the MAE sign. These results demonstrate that the MAE in Fe$_3$GaTe$_2$ is highly sensitive to pressure-driven evolution of the electronic structure near $E_F$, as will be discussed below.

\begin{figure}
    \centering
    \includegraphics[width=1\linewidth]{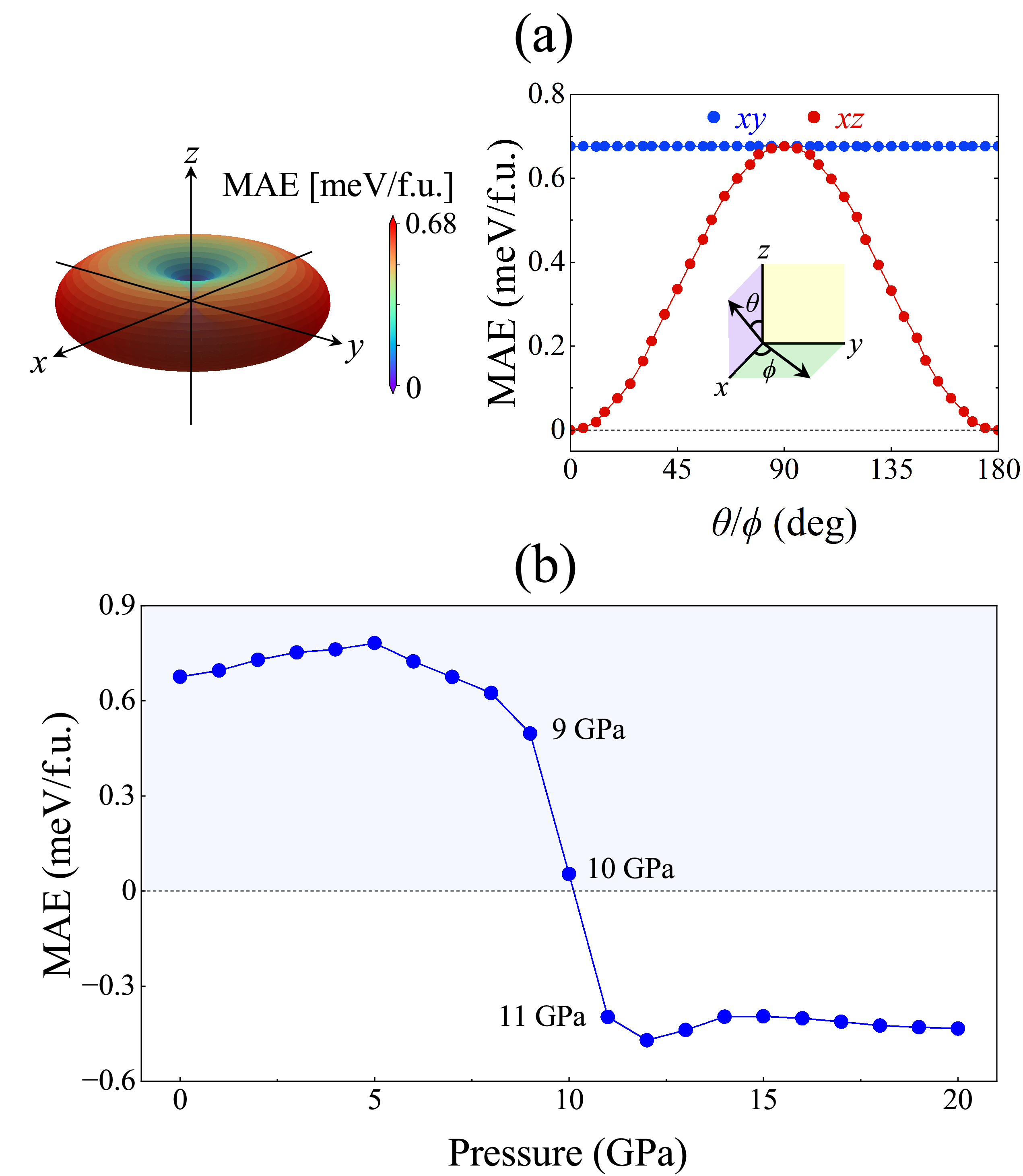}
    \caption{(a) Calculated MAE of Fe$_3$GaTe$_2$ at zero pressure, along with its angular dependence in the $xz$ and $xy$ planes as a function of the angles ${\theta}$ and $\phi$, respectively.
(b) Pressure dependence of the MAE.}
    \label{fig:enter-label}
\end{figure}

To elucidate the microscopic origin of the pressure-induced switching of the magnetic easy axis in Fe$_3$GaTe$_2$, we analyze the atom- and orbital-resolved contributions to the MAE at 9, 10, and 11~GPa, as shown in Fig.~6. This decomposition reveals that the dominant driving forces behind the MAE sign reversal arise from the Fe$_{\rm I}$ and Te atoms. Specifically, the MAE contributions from Fe$_{\rm I}$ (Te) are +0.145 (+0.407), $-$0.106 (+0.252), and $-$0.622 (+0.069)~meV/f.u. at 9, 10, and 11~GPa, respectively, indicating that both atoms together ultimately drive the sign change in the total MAE under pressure. In contrast, the contribution from Fe$_{\rm II}$ increases monotonically with pressure, amounting to 0.129, 0.147, and 0.253~meV/f.u. at the respective pressures. Despite this positive contribution, it is insufficient to counterbalance the increasingly negative contributions from Fe$_{\rm I}$ and Te, resulting in a net reduction and eventual reversal of the total MAE near 10~GPa [see Fig.~5(b)]. Consequently, the magnetic easy axis reorients from out-of-plane to in-plane. These findings indicate that the spin reorientation is primarily governed by pressure-induced modifications in the spin-orbit interaction of Fe$_{\rm I}$ and Te atoms, which exhibit markedly different behavior from that of Fe$_{\rm II}$, as further discussed below.

\begin{figure}
    \centering
    \includegraphics[width=1\linewidth]{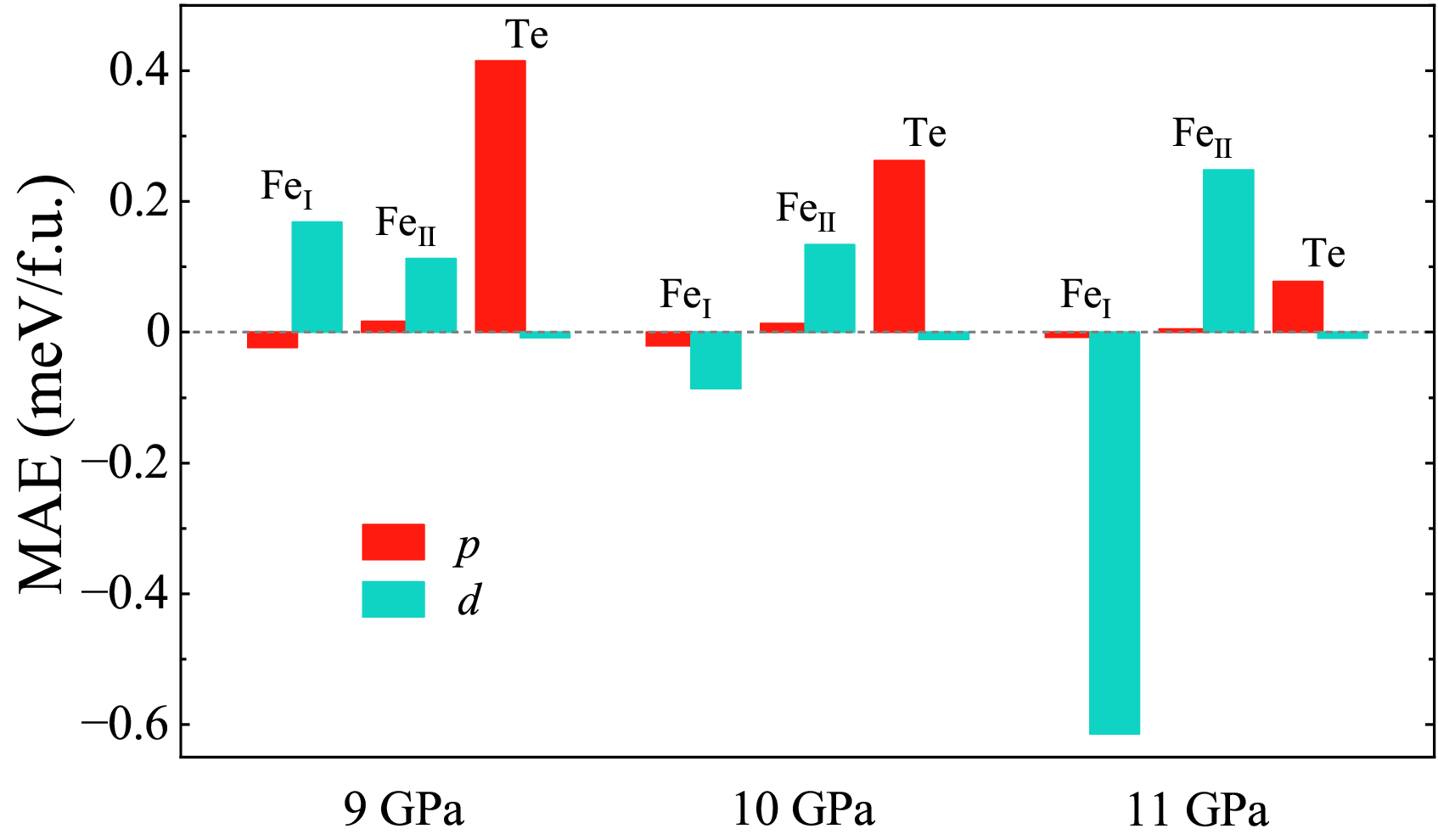}
    \caption{Calculated atom- and orbital-resolved contributions to the MAE in Fe$_3$GaTe$_2$ at 9, 10, and 11 GPa.}
    \label{fig:enter-label}
\end{figure}

Figures~7(a)–7(c) present a detailed orbital-resolved decomposition of the MAE at 9~GPa, illustrating interorbital coupling contributions from Fe$_{\rm I}$, Fe$_{\rm II}$, and Te atoms. For Fe$_{\rm I}$ $d$ orbitals, three major channels are identified: a negative contribution from the ($d_{xy}$, $d_{x^2 - y^2}$) pair, and two positive contributions from the ($d_{xz}$, $d_{yz}$) and ($d_{yz}$, $d_{z^2}$) pairs [Fig.~7(a)]. In contrast, Fe$_{\rm II}$ exhibits moderate positive contributions from the ($d_{xy}$, $d_{x^2 - y^2}$) and ($d_{xz}$, $d_{yz}$) pairs, while the ($d_{yz}$, $d_{z^2}$) pair contributes negatively [Fig.~7(b)]. These SOC interactions yield an overall positive MAE from both Fe sites, with Fe$_{\rm I}$ contributing slightly more, consistent with Fig.~6. For the Te $p$ orbitals, the MAE originates from the ($p_x$, $p_y$) and ($p_y$, $p_z$) channels, both contributing positively [Fig.~7(c)]. The lower panels of Figs.~7(a)–7(c) show the evolution of these interorbital contributions from 9 to 11~GPa. Notably, the MAE contributions from Fe$_{\rm I}$ and Te decrease with pressure and become negative in several channels, indicating a substantial suppression of their magnetic anisotropy. Meanwhile, the contributions from Fe$_{\rm II}$ remain nearly constant or increase slightly, reinforcing the conclusion that the pressure-induced spin reorientation is mainly governed by changes in the MAE contributions from Fe$_{\rm I}$ and Te.

\begin{figure*}
    \centering
    \includegraphics[width=1\linewidth]{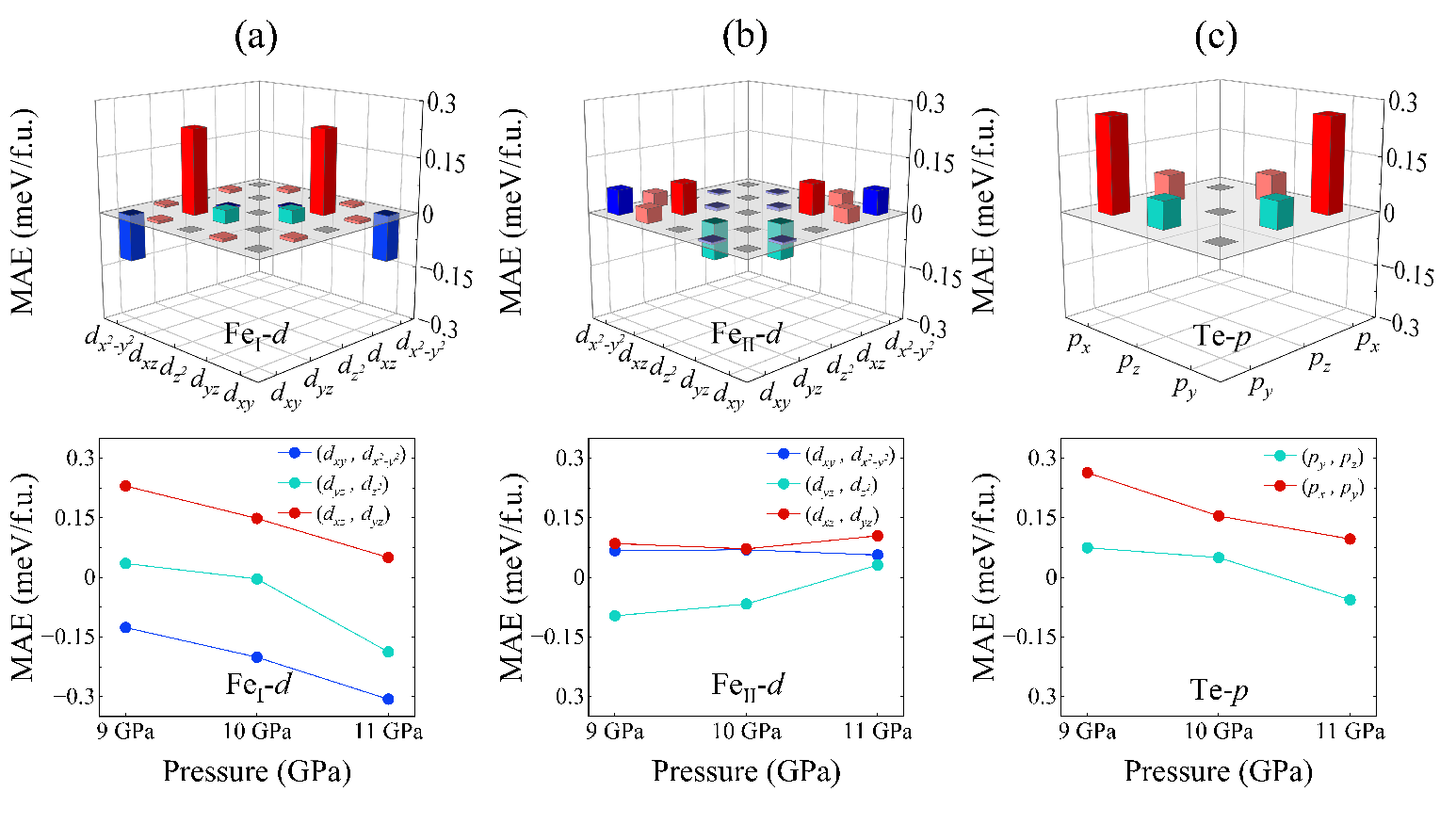}
    \caption{Interorbital contributions to the MAE from (a) Fe$_{\rm I}$ $d$, (b) Fe$_{\rm II}$ $d$, and (c) Te $p$ orbitals at 9~GPa. The lower panels show the evolution of these contributions with increasing pressure from 9 to 11~GPa. Positive (negative) values correspond to a preference for out-of-plane (in-plane) magnetic anisotropy.}
  \label{fig:enter-label}
\end{figure*}

It is worth noting that second-order perturbation theory is formally applicable to systems with discrete and well-separated energy levels, such as insulators or semiconductors~\cite{PRB-EuSn2As2/In2Se3,APL-LaBr2,PRR-MAE}. In metallic systems like Fe$_3$GaTe$_2$, however, its applicability is limited due to strong band hybridization, partial orbital occupancies, and frequent band crossings near $E_F$. These features complicate a strictly quantitative interpretation of SOC-induced magnetic anisotropy. Nevertheless, the contrasting pressure dependence of orbital-resolved MAE contributions from Fe$_{\rm I}$, Fe$_{\rm II}$, and Te atoms can be qualitatively understood by considering their distinct local environments and the effects of anisotropic lattice strain. Under hydrostatic compression, the lattice contracts more significantly along the $c$-axis than within the $ab$ plane [see Fig.~1(c)], resulting in effective uniaxial strain along the out-of-plane direction. This anisotropic deformation modifies the local crystal field and orbital hybridization in a site- and orbital-dependent manner. Fe$_{\rm I}$ and Te atoms, positioned near the vdW gap, are particularly sensitive to interlayer spacing. As pressure reduces this spacing, the spatial extent and overlap of Fe$_{\rm I}$ $d$ and Te $p$ orbitals---especially those with out-of-plane character---are significantly altered. These changes can suppress SOC matrix elements that favor out-of-plane anisotropy or enhance those contributing to in-plane anisotropy, leading to a reduction and eventual sign reversal of their MAE contributions under pressure [see Figs.~6, 7(a), and 7(c)]. In contrast, Fe$_{\rm II}$ atoms are embedded within the interior Fe--Ga layer, where they are coordinated predominantly in-plane and are less directly influenced by interlayer compression. However, pressure can still modify their in-plane crystal field environment and orbital overlap. Such changes may suppress in-plane SOC contributions or amplify those favoring out-of-plane anisotropy, leading to the gradual increase in the Fe$_{\rm II}$ MAE contribution under pressure, as shown in Figs.~6 and 7(b). The divergent pressure responses of Fe$_{\rm I}$/Te and Fe$_{\rm II}$ reflect the interplay between their distinct bonding environments and orbital characteristics. While Fe$_{\rm I}$ and Te respond strongly to interlayer interactions and out-of-plane lattice contraction, Fe$_{\rm II}$, residing in a more symmetric and covalently bonded in-plane network, exhibits a contrasting trend. These findings underscore the importance of anisotropic lattice strain and local orbital geometry in modulating SOC-driven magnetic anisotropy in layered metallic systems. Despite the inherent limitations of perturbation theory in itinerant metallic systems, the present analysis offers a physically consistent and insightful interpretation of the pressure-dependent MAE evolution in Fe$_3$GaTe$_2$.

\section{IV. SUMMARY}

Our first-principles calculations for Fe$_3$GaTe$_2$ revealed a magnetic easy-axis transition from out-of-plane to in-plane at a critical pressure of approximately 10 GPa, in good agreement with experimental observations~\cite{Advanced Science-Fe3GaTe2 pressure}. Around this pressure, pressure-induced electronic band broadening leads to a substantial reduction in magnetization. Notably, the SOC contributions to the MAE exhibit distinct site-dependent behavior. The Fe$_{\rm I}$ and Te atoms near the vdW gap experience enhanced interlayer interactions, which suppress their out-of-plane anisotropy and eventually favor in-plane spin alignment. In contrast, the central Fe$_{\rm II}$ atoms in the Fe–Ga layer show a modest enhancement of out-of-plane preference under pressure; however, this is insufficient to counterbalance the dominant in-plane contributions from Fe$_{\rm I}$ and Te. This contrast in site-resolved SOC responses accounts for the observed easy-axis switching. Our findings highlighted the interplay between electronic band evolution and site-specific SOC effects under pressure in tuning magnetic anisotropy, and demonstrated pressure as a powerful means of controlling spin orientation in layered metallic ferromagnets.

\vspace{0.4cm}

\noindent {\bf ACKNOWLEDGEMENTS} \\

This work was supported by the Natural Science Foundation of Henan (No. 252300421216), the talent Introduction Project in Henan Province (HNGD2025008), the Innovation Program for Quantum Science and Technology (No. 2021ZD0302800), the International Cooperation Project of Science and Technology of Henan Province (No. 242102520029), and the Foundation of Henan Educational Committee (No. 25A140003). J.H.C acknowledges the support from the National Research Foundation of Korea (NRF) grant funded by the Korean Government (Grant No. RS202300218998). The calculations were performed by the KISTI Supercomputing Center through the Strategic Support Program (Program No. KSC-2024-CRE-0055) for the supercomputing application research.

\vspace{0.4cm}

\noindent J. L. and S. L. contributed equally to this work. \\
\noindent $^{*}$ Corresponding authors: wb@henu.edu.cn and cho@henu.edu.cn \\

\noindent {\bf DATA AVAILABILITY} \\

The data that support the findings of this article are not publicly available. The data are available from the authors upon reasonable request.


\begin{thebibliography}{99}
\bibitem{PRESSURE} For the application of hydrostatic pressure, we perform enthalpy minimization with both the cell shape and volume relaxed at each pressure.
\bibitem{MAE-SOC} All MAE calculations were performed with SOC included, which enables the noncollinear formalism in VASP. The magnetization directions were controlled by specifying the global spin quantization axis using the SAXIS tag and appropriately setting the initial MAGMOM values. This setup allowed for a direct comparison between in-plane and out-of-plane orientations. The Tetrahedron method with Bl\"{o}chl corrections was used for the integration without smearing, which is crucial for accurate results in metals. As the $k$-mesh increases to $28 \times 28 \times 7$, the calculated MAE changes by less than a few $\mu$eV/f.u., confirming that the $18 \times 18 \times 5$ k-mesh used in our calculations is sufficiently converged.
\end{thebibliography}
\end{document}